# Nonlinear Sampling and Lebesgue's Integral Sums

Emanuel Gluskin

The Ort Braude and Galilee Sea Academic Colleges, Israel, gluskin@ee.bgu.ac.il

*Abstract* -- We consider *nonlinear, or "event-dependent", sampling*, i.e. such that the sampling instances $\{t_k\}$ depend on the function being sampled. The use of such sampling in the construction of Lebesgue's integral sums is noted and discussed as regards physical measurement and also possible nonlinearity of singular systems. Though the limit of the sums, i.e. Lebesgue's integral, is linear with regard to the function being integrated, *these sums are nonlinear* in the sense of the sampling. A relevant method of frequency detection not using any clock, and using the nonlinear sampling, is considered, in two different versions. The mathematics and the realization arguments essentially complete each other.

## 1. INTRODUCTION

The present investigation is concerned with the possible nonlinearity of the sampling procedure and the associated understanding of Lebesgue's integral scheme. The initial motivation for this research was the logical need to extend the study [1] of singular systems from switching to sampling systems, in order to classify these systems as linear or nonlinear. Below, indices $k$, $p$, $m$, $n$, $s$ and $r$, and also $M$ and $N$ are integers, and 'LTV' means "*linear time variant*".

For both switched and sampling systems, the operation is done at isolated time moments $\mathbf{t}^* = \{t_k\}$ that also are the points of singularity of the *time functions* involved (we mean jump in derivative *of any order* of the function). Symbol $\mathbf{t}^*(.)$, widely used in [1], means that definition of $t_k$ is a point of our concern. System elements switched at $t_k$ are, e.g., $R(t,\mathbf{t}^*)$, $C(t,\mathbf{t}^*)$, and since *state equations* depend on the elements, such notation as $[A(t,\mathbf{t}^*(.))]$ in (1) is natural. Thus, the *switched* "$\mathbf{t}^*(.)$-systems" can be classified [1] using the usual equations ($\mathbf{x}(t) = \{x_p(t)\}$, $\mathbf{u}(t) = \{u_m(t)\}$)

$$d\mathbf{x}/dt = [A(t,\mathbf{t}^*(.))]\mathbf{x}(t) + [B(t,\mathbf{t}^*(.))]\mathbf{u}(t) \qquad (1)$$

in which the linear (LTV) case is that of $\mathbf{t}^*(t)$ (i.e. all $\{t_k\}$ are prescribed), and the nonlinear (NL) case is that of $\mathbf{t}^*(\mathbf{x})$ (when, e.g., some of the $t_k$ are some level-crossings of some of the $x_p(t)$). This means, in particular, that the LTV and NL cases are given in (1) *in the same structural terms*.

A more unexpected nonlinearity of the switched systems is that of $\mathbf{t}^*(\mathbf{u})$, when the switching instances are directly influenced by the input. The theorem of [2], not dealing with the structural aspect, says that if a switching occurs because of increase, starting from zero, of the scaling factor of the input $Uf(t)$, then the system is nonlinear. This case is more close to the situation with the sampling systems.

## 2. NONLINEAR SAMPLING

In the simplest case, sampling is nonlinear if the sampling instants depend on the function $f(t)$ being sampled. See Fig.1. To explain this case, let us start from the fact that in the terms of the sampled values, or *evaluation*, the basic addition procedure, included in the axioms of linear space is

$$(f_1 + f_2)(t^*) = f_1(t^*) + f_2(t^*) \qquad (2)$$

which is just $T[f_1+f_2]=T[f_1]+T[f_2]$ for the sampling operator $T: t \to t^*$ (i.e. a convolution with $\delta$-function) *when this operator is linear*, -- see the argument below.

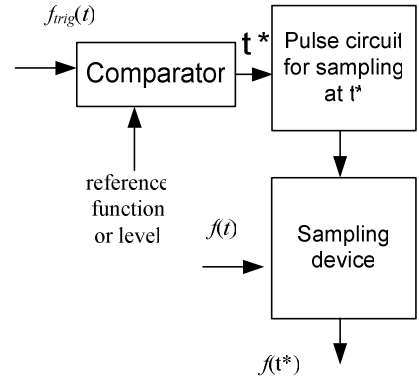

Fig. 1: The basic scheme of electronic sampling. If $f_{trig}(t)$ and the function $f(t)$ to be sampled are analytically connected, or, most simply, the switching instants $\mathbf{t}^*$ depend on $f(t)$, then sampling becomes nonlinear.

Equation (2) allows us to make the needed observation and formulate the point that connects circuitry (hardware) matters with mathematics. In this sense, this simple equality replaces here, at the start, equation (1) related to the switching systems.

Let us see in the context of (2) our '$T$' as '$T_{t^*}$', i.e. make it possible to take different $t^*$ for the different terms. Since it is obviously necessary for the usual linear sampling that $t^*$ be the same in each term, *the linear case requires the sampling instants of an electronic sampler to be defined independently of the function being sampled*, and understanding this point just opens the way for a definition of *nonlinear* sampling. If $f_{trig}$ and the sampling instants are *dependent* on the function being sampled, then (2) is non-checkable by the (then, nonlinear) sampling system having the time-functions at its inputs; the comparator that precedes the sampler and triggers it will define (apart from some special cases) different sampling instants for the distinct terms. Letting $f_{trig}(t)$ and $f(t)$ be "rigidly connected" and the comparator's reference be constant, we even cause all terms in (2) to have the same value, and if this value is nonzero, (2) is incorrect.

Also for the *scaling factor* in the function $Uf(t)$ to be sampled, the evaluation of the type $Uf(t^*(U))$ is nonlinear by $U$. Sampling defined by levelcrossings, generally introduces nonlinearity. Sampling nonlinearity also can be provided by a special ("additional", see Sect. 4.1) inclusion of $t_k$ into the function to be sampled.

Compare the noted cases with the "**t***(.)-outlook" of (1)".

### 3. LEBESGUE'S INTEGRAL SUMS AND THE "EVENT-DEPENDENT" SAMPLING

There is a noticeable similarity of the nonlinear sampling and the construction of Lebesgue's integral sums in which the small steps are made not along the axis of the argument, but along the *f*-axis (Fig. 2).

Though, the limit of the sums, i.e. Lebesgue's integral, is linear with regard to the function being integrated, one sees that the sums are nonlinear in the sense of the nonlinear sampling.

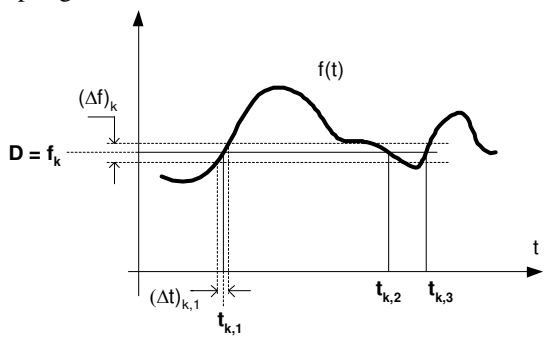

Fig. 2. Construction of Lebesgue's integral. Using technical language, no time clock is given, and the intervals $(\Delta t)_{k,m}$ and the sampling instances $t_{k,m}$ are not taken independently as in Riemann's integral; they depend on $f(t)$. This leads to nonlinearity of the approximating sum (7), or (8), that includes terms of the type $f(t_k)(\Delta t)_k$.

The points $t_k$ are obtained as the instants of the crossings of the level $D$ by the input function. That is, in the realization we compare, using electronic comparators, the values of the input function with the "test-levels" $D$. Changing $D$, e.g., by giving quantization, i.e. using an equal-distant set $D_k = k\Delta_D$ ($\Delta_D$ is small and fixed/constant below), we obtain different $t_k$.

The unit of the mathematics with the realization, can be schematically illustrated by the block-scheme in Fig.3, which describes, in terms of the hardware, the *precisely realized* mathematical identity:

$$f(t^*) = D \quad \text{(correct, by the definition of } t^*\text{)} \quad (3)$$

As is seen from this figure, the instant $t^*$ (the "output" of this logical scheme) is defined by the crossing by $f(t)$ of the level $D$ (an "input" of the scheme) i.e. as $f^{-1}(D)$, and $f_k = D$ is simultaneously defined as $f(t^*)$. Mathematically, this is the identity $D = f(f^{-1}(D))$, though not a universal one since $f^{-1}(.)$ can be only *locally* defined.

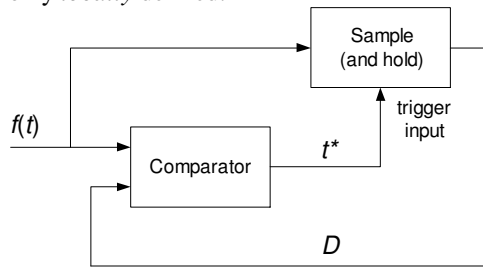

Fig. 3: Realization of (2). For given (in a proper range) $D$, $t^*$ is defined by $f(t)$. Note that for Lebesgue's sampling, this figure replaces Fig.1.

Realizing this scheme and studying its stability should be relevant to modeling the finite sums as (8) approximating Lebesgue's integral. See Section 4.2.

Since $f(t)$ both defines $t^*$ and is sampled at it, Lebesgue's sampling is nonlinear by definition. That this yields nonlinearity of the integral sum is also seen directly. Consider the function possessing the derivatives needed for writing (see Fig. 2 where $(\Delta t)_{k,m}$ are defined):

$$(\Delta f)_k = \frac{df}{dt}(t_{k,m})(\Delta t)_{k,m} + \frac{1}{2}\frac{d^2 f}{dt^2}(t_{k,m})(\Delta t)^2_{k,m} + \ldots . \quad (4)$$

If $(df/dt)(t_{k,m}) \neq 0$, then for some sufficiently small $(\Delta t)_{k,m}$, we have, for each $m$,

$$(\Delta t)_{k,m} \approx \frac{(\Delta f)_k}{(df/dt)(t_{k,m})}, \quad (5)$$

at the level $f_k = D = f(t_{k,m})$, which defines $t_{k,m}$, $\forall m$. Since $sign[\Delta f] = sign[df/dt]$, all $\Delta t$ given by (5) are positive.

If $(df/dt)(t_{k,m})$ is zero, but the second derivative at $t_{k,m}$ is nonzero, then, taking from (4) positive value of $(\Delta t)_{k,m}$, we have, for some sufficiently small $(\Delta t)_{k,m}$, that

$$(\Delta t)_{k,m} \approx \sqrt{\frac{2(\Delta f)_k}{(d^2 f/dt^2)(t_{k,m})}}, \quad (6)$$

and so on. As the point, we always have $(\Delta t)_{k,m}$ (and also $t_{k,m}$) dependent on $f$. Thus the "blocked" approximating sum, related to the whole range of $f(t)$, which uses the *measures*

$$meas(f_k, \Delta f_k) = (\Delta t)_{k,1} + (\Delta t)_{k,2} + \ldots, \text{ for each level } D = f_k,$$

leads to the *nonlinear by f* Lebesgue's sum

$$\sum_k f_k \, meas(f_k, \Delta f_k) = \sum_k f_k \left(\sum_m \Delta t_{k,m}\right). \quad (7)$$

When $f(t)$ is an unknown input process, and the integral is determined on line by electronic equipment, we cannot know how the summation by $m$ will arise, and have to see (7) as

$$\sum_{n=1} f_n \cdot (\Delta t)_n. \quad (8)$$

where *n* labels, sequentially in time, *all* the time intervals. Riemann's integral sum is also thus written, however Riemann's sum has independent values "$\Delta t$", and thus is linear with respect to the discretely presented input, just as the precise integral is with respect to the precise function.

Let us consider now, following [9], a qualitative possibility of frequency measurement associated with an unusual sampling, and then try to use Lebesque's approximating sum, as an alternative possibility. Event-dependent sampling is also considered (from different positions) in recent works [3-8], while in [3] some interesting old references, starting from 1965, are given.

### 4. THE $\Psi$-TRANSFORM OF A SIGNAL WHOSE ZERO-CROSSINGS ARE THE "EVENTS", AND THE "CONVERSING" SIGNAL-RECEPTION

Nonlinear sampling can be also illustrated by the nonlinear "$\psi$-transform", introduced in earlier work [9]:

$$f(t) \to \psi(t): \quad \psi(t) = \int_0^t f^2(\lambda)\, sign[f(\lambda)]\, d\lambda \qquad (9)$$

where $f(t)$ is the input function (process) to be analyzed.

In the method of [9], $\psi(t)$ is sampled (Fig. 4) at the zerocrossings $t_k$ of $f(t)$, and an estimation of some basic spectral parameters of a stationary signal, *not using any "clock"*, is based on these sampled values.

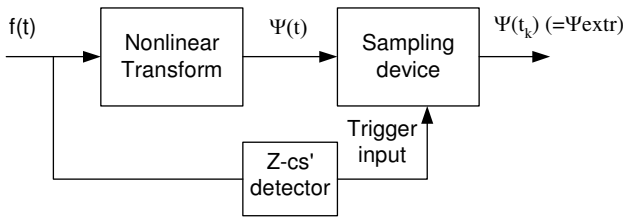

Fig. 4. The schematic for obtaining $\{\psi(t_k)\}$, where $t_k$ are the zerocrossings of $f(t)$ (or of $f(t) - \langle f(t)\rangle$), these parameters to be used for estimating the average period/frequency of the process $f(t)$. (Note that it appears [9] that $\{\psi(t_k)\}$ are the extreme values of $\psi(t)$, and are of alternating sign.)

Observe that $t_k$ are involved in our procedure *twice*, -- in the definition of the rectangular-wave function $\varsigma(t) \equiv sign[f(t)]$ appearing in the integrand, and in $\psi(t_k)$. (For any certain $f(t)$, $\psi(t)$ depends on $\{t_k\}$ *explicitly*, [10] for such details; and thus any particular $\psi(t_{k'})$ depends on all the other $t_k$, $k < N$, where $N$ is the number of the zero-crossings of $f(t)$, which arrived, -- our basic parameter.)

Compared to Fig.1: "$f_{trig}(t)$" there, is $f(t)$ here, the "reference level" is zero, "$f$" there, is our $\psi(t)$, i.e. $f_{trig}$ is included in *f*.

### 4.1. The measurement of the average frequency (period) of a signal/process without giving any clock

Function $\psi(t)$ given by (9) may be assumed to be limited, for the period of the processing, and thus can be generated in an analog, i.e. quick, manner. However, the integration of $f^2(t)$, leading to the (very important below) "energy function"

$$E(t) = \int_0^t f^2(\lambda)\, d\lambda \quad , \qquad (10)$$

is all the time increasing, and thus is irrelevant for *analog* performance where it would be the energy of real physical elements. One can think about *numerical integration*, but still we see $E(t)$ as the usual Riemann's integral this is also rejected, because digital evaluation of such an integral requires a prescribed rate of sampling of $f^2$, i.e. giving a time or frequency unit of the clock that we wish to generally avoid here.

Comment 1: Let *the sender of the signal* define the clock unit, $\delta t$, of time-intervals measurement, and consider a communicating robot, exploring an environment, which has to decide whether or not this (some) unexpectedly received signal is of interest to it and should be studied. It could be that a high-precision detailed treatment of the received signal is not needed at this checking stage, i.e. we can use our simple procedure (see also [11]) and even can limit N, by, say, the small "$N_{needed}$" = $10^3$, or $10^4$, allowing the robot just understand, e.g., whether or not it is Hebrew speech being transmitted. The period of the processing is a function of $N_{needed}$ and $\delta t$, but is *unknown a priori*, because $\delta t$ is not given. That the information is treated at the rate it is received ($\sim \delta t$), is quite similar to usual human conversation, and if the information is supplied slowly, then the whole "conversation" is slow, but the "on-line understanding" of whether or not the information is relevant should be quick, because of the analog procedures involved in the processing. The reception without clock might be named "conversing reception". □

However, we can use $\{\psi(t_k)\}$ instead of $E(t)$, and thus avoid using any special frequency unit, determining nevertheless, the average frequency of the process $f(t)$. For this, we shall also need the "average power" of the signal,

$$P(t) = <f^2>_t = \frac{1}{t}\int_0^t f^2(\lambda)\, d\lambda = E(t)/t, \qquad (11)$$

which is a limited function, and thus *can be obtained on line by the (analog) low-pass filtering* of $f^2(t)$.

Purposed to estimate the *average* period $T_a$ of a compact spectrum of the signal, we act as follows. Assuming that on average there should be two zero-crossings per $T_a$ (as it is for a sinusoid), we define $T_a$ in agreement with the equality $N \approx 2t/T_a$, i.e. we assume that the following average (seen as either an average by $t$, or by $N$, for $N \gg 1$) exists:

$$T_a = 2\langle t/N \rangle. \qquad (12)$$

Since, in view of (10) and (11), $t = E/P$, (12) means that

$$T_a \approx 2\frac{t}{N} = 2\frac{E(t)}{NP} \quad . \tag{13}$$

We now perform the transformation of $E(t)$ in which the main part/term is associated with the main interval $(t_1, t_N)$, $N \gg 1$, of the observation (treatment) of the signal:

$$E(t) = \int_0^t f^2(\lambda)d\lambda = \int_0^{t_1} f^2(t)dt + \int_{t_1}^{t_N} f^2(t)dt + \int_{t_N}^t f^2(\lambda)d\lambda$$

$$= \int_{t_1}^{t_2} f^2(t)dt + \int_{t_2}^{t_3} f^2(t)dt + \ldots + \int_0^{t_1} f^2(t)dt + \int_{t_N}^t f^2(t)dt$$

$$= \int_{t_1}^{t_2} f^2(t)\operatorname{sign} f\, dt - \int_{t_2}^{t_3} f^2(t)\operatorname{sign} f\, dt + \int_{t_3}^{t_4} f^2(t)\operatorname{sign} f\, dt - + \ldots$$

$$= [\psi(t_2) - \psi(t_1)] - [\psi(t_3) - \psi(t_2)] + [\psi(t_4) - \psi(t_3)] - + \ldots$$

$$= -\psi(t_1) + 2\psi(t_2) - 2\psi(t_3) + 2\psi(t_4) + \ldots$$

$$= \psi(t_1) - \psi(t_N) + 2 \sum_1^{N/2} \left[\psi(t_{2p}) - \psi(t_{2p-1})\right] +$$

$$+ \int_0^{t_1} f^2(t)dt + \int_{t_N}^t f^2(\lambda)d\lambda. \tag{14}$$

Since as $t \to \infty$, i.e. for $N \gg 1$, $E(t) \to \infty$, the several (all finite) terms in (14) which are not bracketed are relatively small compared to the value of the sum, we can write, for large $N$, (14) as the sum of the (positive) bracketed values

$$E(t) \approx 2 \sum_{p=1}^{N/2} \left[\psi(t_{2p}) - \psi(t_{2p-1})\right] \quad . \tag{15}$$

(See also the inversion of this relation in [9].)

According to (13) and (15), we have now:

$$T_a = \lim_{t \to \infty} \frac{2E(t)}{N(t)P(t)} =$$

$$= \lim_{t \to \infty} \frac{4 \sum_{p=1}^{N/2} \left[\psi(t_{2p}) - \psi(t_{2p-1})\right]}{NP(t_N)} = ,$$

$$= 2 \lim_{N \to \infty} \frac{\langle \psi(t_{k+1}) - \psi(t_k) \rangle_N}{P(t_N)} \tag{16}$$

averaging by $N$ the differences between close maxima and minima of $\psi(t)$, which are the $\psi(t_k)$. In the periodic case when $f(t+T/2) = -f(t)$, we simply obtain $T_a = T = 4|\psi(t_1)|/P$.

Note that, finally, the number of zero-crossings counted "on line", and not $t$, is the central parameter.

*No clock, i.e. special unit of time, is used here. It is just required that some limitation on the spectrum of f(t) be roughly assumed* (*known*), *in order for the integrator, the sampler, and other involved devices to work. The latter is however, a general requirement of any technical device, including one that can work in a very wide frequency range, and this is not intended to specify a unit of measurement.*

Comment 2: That we do not give any special unit of measurement in this procedure does not mean, of course, that we have here a problem with the basic physics units. The value of $T_a$ defined by (16) appears in the usual time units *because of the physical analog processes involved*. Integral (9) that introduces $t$ in (16) is, in fact, a solution of a differential equation that describes a continuous process in a circuit/integrator, and in the analog generation, $\psi$ automatically (physically) appears in the units $[f]^2[t] = [P][t]$. (Compare this with charging a capacitor, by current $i$ up to a measured charge $q$, when the period of the process can be estimated as $q/i$.) □

The described procedure is the basic step of the more general procedure of *spectrum analysis* of a function, not using any clock, based on the sampling of the function's $\psi$-transform at its zero-crossings. In [11], this step is proposed to be applied not only to $f(t)$ appearing in the input channel, but also to the time processes obtained from $f(t)$ by some sequential reductions of the spectrum of $f(t)$, by means of channel splitting. At the inputs of the new channels, the cutoff (splitting) frequencies of these reductions will be defined on-line by the "$T_a$", found in the previous processing step. Since we use $P$ for finding $T_a$, at each application of the "basic step", i.e. in each channel, we have the pairs $\{P, \omega_a = 2\pi/T_a\}$ that are the points on the shape of the power spectrum $S(\omega)$ of $f(t)$.

This generalization is relevant also to the other "basic step", described below.

### 4.2. Estimation of $E(t)$ in (13) as Lebesgue's integral sum

Returning to the possibility of directly numerically estimating $E(t)$, which was previously rejected because of a clock needed for sampling $f^2(t)$ in Riemann's version/construction of integral (10), let us note that if the integration of $f^2(t)$ could be done according to Lebesgue's scheme, i.e. without a clock, then, in principle, we could use $E(t)$ and not $\psi(t)$, for the determination of $T_a$, i.e. (13) would be sufficient and (14)-(16) not needed.

Thus, returning to $\{t_n\}$ defined by the quantization condition $f(t_n) = n\Delta_D$ (see Fig. 2, noting that index $k$ is reserved in Section 4.1 for the zeros of $f(t)$, which subdivide the time axis in a much rougher scale; the same relates to '$N$' that should not be mixed with '$M$' below), we have, according to (8), that still $n$ of the level being crossed is increased for $n < M_1$, with some $M_1$, then for this interval of monotony (monotonicity) of $f(t)$

$$E(t_M) \approx \sum_{n=1}^M (f_n)^2 \cdot (\Delta t)_n = \Delta_D^2 \sum_{n=1}^M n^2 (\Delta t)_n$$

$$= \Delta_D^2 \sum_{n=1}^M n^2 (t_{n+1} - t_n), \quad M < M_1 \tag{17}$$

and when, later, $n$ is decreased, then

$$E(t_M) \approx \Delta_D^2 \sum_{n=1}^{M_1} n^2 (t_{n+1} - t_n) + \Delta_D^2 \sum_{s=0}^{M-M_1} (M_1 - s)^2 (t_{M_1+s+1} - t_{M_1+s}), \quad (18)$$

where $M_1 < M < M_2$, with some $M_2$, and so on, with more such sums for the oscillating function $f(t)$, and $n$ (or $n(s)$) sometimes increasing, and sometimes decreasing.

The involved $t_n = t_n(f, \Delta_D)$ are meant to be found, as time passes, using the electronic scheme of the type of Fig.3.

Since $\Delta_D \sim$ "$f_{max}$"/$M_1$, equalities (17), (18), ... lead to $T_a$ in (13) via some specific averaging in $E(t)$ of the small time intervals associated with the quantization. That $P$ in (13) is measured in analog way avoids use of clock here too.

Thus, in principle, (13) can be realized *without any clock* either by replacement of $E(t)$ using the $\psi$-transform as in (15), or by the more direct approximation of $E(t)$ by Lebesgue's integral sum. In both cases, nonlinear sampling is involved and the data (apart from that related to the analog generating $\psi(t)$ and $P(t)$) accumulated (stored) numerically.

It can be suggested that electronics specialists should try to develop realizations of Lebesgue's approximating sums.

5. CONCLUDING REMARKS

The distinction between Lebesgue's and Riemann's schemes is not only in the possibility to integrate, e.g., the Dirichlet's function (that equals 1 for a rational value of the argument, and 0 for irrational value [12]), but also in the nonlinear, "event-dependent", sampling of physical functions. It is assumed therefore (see also [3-8]) that Lebesgue's approximating sums can be useful in signal processing, which gives an "engineering justification" to Lebesgue's scheme, and can motivate associated electronic design.

The comparator involved in the Lebesgue's scheme works in terms of the relations '>' or '<' that belongs to comparison of *real, and not complex* numbers. Thus, Lebesgue's scheme dictates reality of the input function $f(t)$, and it becomes clear why the classical situation is that [13] Lebesgue's integral is applied to real, and not complex analysis.

The reality of $\delta$-function inherently associated with sampling operation, is also of the same nature.

The role of the quantization in the Lebesgue's sampling is not quite usual. If (see Fig.2) we were measuring $f(t)$ in the units $\Delta_D$ *at prescribed time instances $t_r$*, then there would be some "quantization noise" defined by the errors $\{f(t_r) - s_r \Delta_D\}$, $\{s_r\} \subset \mathbb{Z}$, of order $\Delta_D/2$. In the Lebesgue's sampling, we *wait* until the levelcrossings occur, meaning $f(t_r) = s_r \Delta_D$, $\forall r$, which, in principle, eliminates this noise. Thus, Lebesgue's sampling might be also named *noise-free quantization*.

Contrary to the switched systems described by (1), classification of the sampling systems cannot be done in the detailed structural terms, only in terms of operations with time functions, and thus while *linear* switched system (structure) *must be* LTV, sampling can be "simply" linear.

ACKNOWLEDGMENT

I am grateful to Steven Shnider for helpful comments regarding structure of the work, to Anthony J. Weiss for good style advices and a question that led me to Comment 2, and to Shmuel Miller and Joos Vandewalle for kind attention and encouragement.